# Repeatable Room Temperature Negative Differential Conductance in GaN/AlN Resonant Tunneling Diodes

Jimy Encomendero[1], Faiza Afroz Faria[2], S.M. Islam[1], Vladimir Protasenko[1],
Sergei Rouvimov[2], Patrick Fay[2], Debdeep Jena[1,3] and Huili Grace Xing[1,3]

[1]*School of Electrical and Computer Engineering, Cornell University, Ithaca, NY 14853 USA*

[2]*Deparment of Electrical Engineering, University of Notre Dame, Notre Dame, IN 46556 USA*

[3]*Department of Materials Science and Engineering, Cornell University, Ithaca, NY 14853 USA*

Double barrier GaN/AlN resonant tunneling heterostructures have been grown by molecular beam epitaxy on the (0001) plane of commercially available bulk GaN substrates. Resonant tunneling diodes were fabricated; room temperature current-voltage measurements reveal the presence of a negative differential conductance region under forward bias with peak current densities of ~6.4 kA/cm$^2$ and a peak to valley current ratio of ~1.3. Reverse bias operation presents a characteristic turn-on threshold voltage intimately linked to the polarization fields present in the heterostructure. An analytic electrostatic model is developed to capture the unique features of polar-heterostructure-based resonant tunneling diodes; both the resonant and threshold voltages are derived as a function of the design parameters and polarization fields. Subsequent measurements confirm the repeatability of the negative conductance and demonstrate that III-nitride tunneling heterostructures are capable of robust resonant transport at room temperature.

## Introduction

Resonant tunneling of electrons in III-V semiconductors has undergone extensive studies since Tsu and Esaki theoretically investigated transport across multi-barrier heterostructures with periods comparable to the electron's wavelength[1]. Quantum confinement introduced by the barriers results in a localized and discrete electronic spectrum, which can be tuned to bring the quasi-bound levels in energy resonance with any adjacent reservoir of electrons. The resonant transport regime has been exploited to design highly efficient injectors of electrons into the upper lasing level of terahertz (THz) quantum cascade lasers (QCLs)[2]. On the other hand, the out-of-resonance regime, identified by the onset of negative differential conductance (NDC), has been harnessed to realize high frequency resonant tunneling diode (RTD) oscillators[3]. Despite the steady progress in output power and frequency of operation, RTD oscillators are yet to be demonstrated at the miliwatt output power level for >1 THz, which is required for most practical applications[4]. Meanwhile, THz-QCLs fabricated with the well-developed material systems AlGaAs/GasAs and InGaAs/InAlAs have not yet achieved room temperature operation and their lasing frequencies are limited to <5 THz mainly due to Reststrahlen absorption[5]. In this scenario, the GaN/AlGaN material system has emerged as an attractive alternative to realize intersubband emitters within a wide range of frequencies due to the large conduction band energy offset of ~1.75 eV between GaN and AlN[6]. In addition, the high longitudinal optical (LO) phonon energy of the III-nitride materials, ~92 meV in GaN, is expected to prevent the depopulation of the upper lasing level, thus raising hopes for room temperature operation of nitride THz-QCLs[7].

Double barrier GaN/AlGaN RTDs, being the simplest device to study resonant transport, have been under scrutiny during the last decades with moderate success[8–23]. Experiments on AlN-barrier RTDs grown on sapphire templates have shown a region of NDC under the forward bias of operation, with the bottom contact layer as the reference of the applied voltage. However, the measurements are also characterized by a lack of repeatability and a clear hysteretic behavior which prevents the measurement of NDC in the downward sweep[10,18,21,22]. It has been suggested that the high density of defects present in GaN films grown on sapphire (~$10^9$ cm$^{-2}$) can act as electron traps, leading to self-charging and preventing coherent transport of carriers. Low temperature (LT) superlattices[8] and lateral epitaxial overgrown (LEO) films[16] have been also employed in order to reduce the density of defects during growth; however the reported negative conductance still degrades with consecutive measurements[17]. III-nitride RTDs with low Al-composition AlGaN-barriers have been also studied at cryogenic temperatures[19]. These devices were grown on bulk-GaN substrates with low dislocation densities (~$5\times10^6$ cm$^{-2}$) and small mesas of 16 μm$^2$ were defined to reduce the number of defects per device. Repeatable NDC features were observed only at cryogenic temperatures, below 110 K for a diode with 18%-AlGaN barriers and below 130 K for a diode with 35%-AlGaN barriers[19,20,23]. In this work, GaN/AlN RTDs were grown by molecular beam epitaxy (MBE) and fabricated using conventional lithography techniques. Current-voltage (I-V) measurements show a clear resonant peak and a stable and repeatable NDC region of operation under forward bias, thus demonstrating that III-nitride heterostructures are capable of robust resonant tunneling transport at room temperature. The repeatable operation has helped uncover several unique features in the tunneling spectrum that originate from the spontaneous and piezoelectric polarization in these heterostructures.

**Experiment**

A symmetric double barrier GaN/AlN heterostructure was grown by MBE on the c-plane of commercially available n-type bulk GaN substrates with a nominal dislocation density ~$5\times10^4$ cm$^{-2}$. Metal rich growth conditions were maintained during the whole process at a substrate temperature of ~700 ºC and a nitrogen plasma power of ~200 W. A growth rate of ~5 seconds per monolayer (ML) was estimated by means of post-growth structural characterization methods. The active region comprises a ~3 nm GaN quantum well with ~2 nm AlN barriers and un-intentionally doped (UID) spacers of ~2 nm next to both barriers as indicated in Fig. 1(a). The emitter and collector regions consist of ~100 nm n-type GaN layers with a silicon doping level of ~$7\times10^{18}$ cm$^{-3}$. A highly Si-doped (~$8\times10^{19}$ cm$^{-3}$) GaN cap layer completes the epitaxial heterostructure allowing for the formation of an ohmic contact at the collector metal-semiconductor junction.

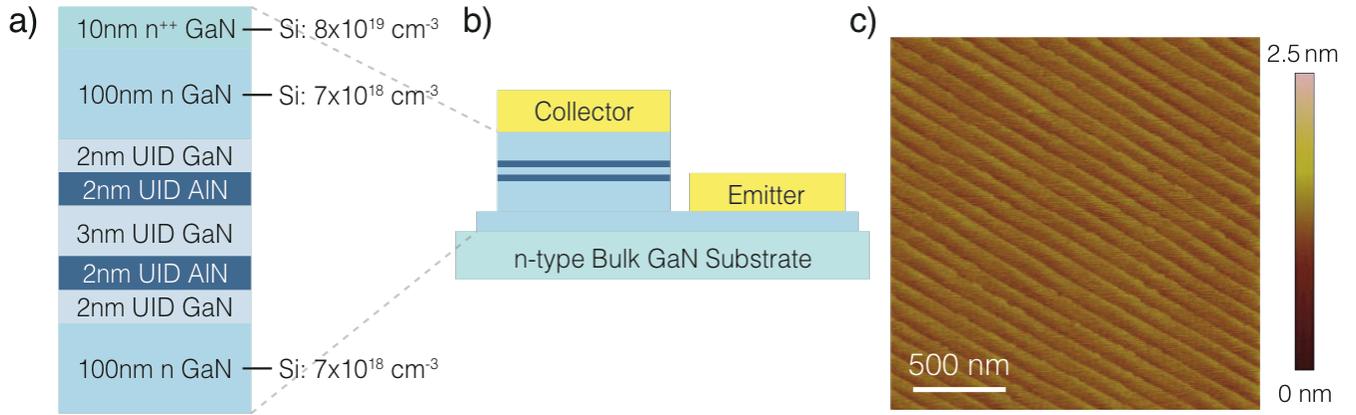

FIG. 1. (a) Schematic of the layer thicknesses and doping concentrations of the GaN/AlN double barrier heterostructure grown by MBE on a bulk GaN substrate. (b) Cross section schematic diagram of a fabricated resonant tunneling diode with the collector and emitter metal contacts. (c) Surface morphology of the as-grown RTD heterostructure with an rms roughness of ~0.146 nm over a scanned area of 2×2μm$^2$

MBE growth of GaN films has been studied extensively in recent years, showing that smooth surfaces can be generated under excess metal growth conditions. In this regime, a Ga layer of ~ 2.5 ML is present at the growth interface favoring layer-by-layer incorporation of GaN[24,25]. Thus, Ga-rich conditions have been employed during the growth of the emitter and collector n-GaN layers as well as during the growth of the quantum well. However, using excess Al metal during the growth of AlN barriers leads to thicker AlN layers since the excess Al readily incorporates in the crystal in preference to Ga[26]. An alternative approach to promote layer-by-layer growth of AlN layers exploits the surfactant effects of excess Ga, which decreases the surface energy without getting incorporated into the AlN films[25]. In this work, excess Ga has been employed throughout the epitaxial growth in order to enable layer-by-layer growth and to minimize defect generation. Furthermore, crystal defects stemming from strain relaxation processes should be suppressed since the AlN barriers are thinner than the experimentally identified critical thickness (~5-7 nm) for AlN films pseudomorphically grown on GaN[27,28] Atomic force microscopy (AFM) was employed to image the surface of the as-grown sample, revealing a topology composed by 2 ML atomic steps with a root mean square (rms) roughness of ~0.146 nm over an area of 2×2 μm$^2$ (Fig. 1(c)).

The fabrication of the RTDs was carried out using conventional contact photolithography, electron beam evaporation and reactive ion etching techniques. The collector contact was defined by evaporating the Ti/Al/Au/Ni metal stack. Mesas with areas between 6-48 μm$^2$ were defined using a self-aligned process in which the structure was etched down to the emitter n-GaN layer by reactive ions of the Ar/Cl$_2$/BCl$_3$ gas mixture. Finally, the emitter Ti/Al/Au metal stack was deposited to obtain the structure depicted in Fig. 1(b). Transmission line measurement (TLM) structures were also defined; a specific contact resistance of ~30 Ω−μm$^2$ was measured in the collector contacts and the emitter contact resistance was estimated in the same order of magnitude. Thus, under injection currents on the order of ~10 kA/cm$^2$, a voltage drop of ~3 mV per contact is expected, which is negligible compared to the total voltage used to bias the diodes.

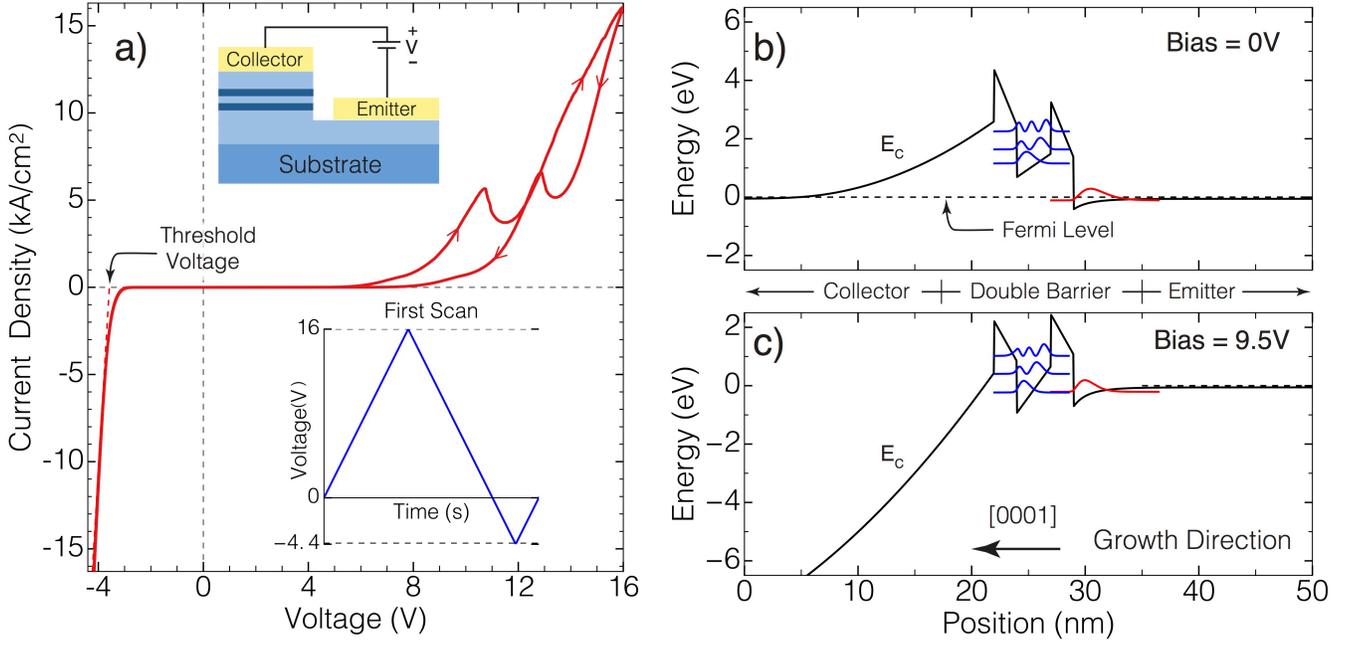

FIG. 2. (a) GaN/AlN RTDs current-voltage characteristics at room temperature during the first bias sweep carried out by applying a closed-loop bias scan (shown in the bottom inset of (a)). During the upward sweep, a resonant peak is measured at ~10.4 V with a peak-to-valley current ratio of ~1.5 and a peak current of density of ~5.7 kA/cm². A hysteresis loop is measured during the downward sweep with the resonant peak shifted towards a higher voltage (~12.9 V) (b). Band diagram under equilibrium conditions showing the accumulation subband in the emitter and the resonant states in the quantum well. (c) Band diagram under resonant conditions enabled by the energy alignment between the accumulation subband next to the emitter AlN barrier and the ground level inside the quantum well.

## Results

Current-voltage (I-V) characteristics were measured with a semiconductor parameter analyzer at room temperature applying voltage sweeps starting at 0 V up to 16 V, and then back to 0 V (double sweep). Immediately after, a reverse bias double sweep with a minimum voltage of −4.4 V was also performed, completing a closed loop scan (inset of Fig. 2(a)). The polarity of the diode bias is that of the voltage applied to the collector side, having the emitter as reference as shown in the top inset of Fig. 2(a). At forward biases below 5 V, current densities below 10 A/cm² were measured. These low injection currents are a result of the combination of a large GaN/AlN conduction band discontinuity (~1.75 eV) and polarization fields that effectively blocks carriers thus preventing transport from the emitter to the collector side. The equilibrium band diagram shown in Fig. 2(b)—which has been calculated using the nominal thicknesses of the layers comprising the heterostructure—illustrates this blocking effect.

For voltages larger than 5 V currents larger than 10 A/cm² were recorded. During the first forward scan, a resonant peak is measured in the upward sweep at ~10.7 V with a peak current density of ~5.7 kA/cm² as shown in Fig. 2(a). The adjacent NDC region extends up to ~11.5 V where the valley current is measured at the level of ~3.7 kA/cm². For higher biases (>12 V), the current increases exhibiting an exponential trend resembling the characteristics of RTDs fabricated with well-developed materials systems. The resonant peak is attributed to the enhancement of transmission probability for carriers populating the accumulation subband adjacent to the emitter AlN barrier. The RTD band diagram corresponding to this resonant condition is shown in Fig 2(c).

During the downward sweep of the first closed-loop scan, a hysteretic behavior is measured with the resonant bias shifted towards a larger voltage (~12.9 V). The hysteresis loop and shift in resonant bias, point to the presence of trap charging mechanisms, which modify the electrostatics profile of the heterostructure. A detailed study of the dynamics, activation energy and location of the traps is beyond of the scope of this work. Nevertheless, despite the presence of these crystal imperfections, resonant transport of electrons remains as the main conduction mechanism at room temperature evidenced by the presence of the resonant peak and NDC region in both upwards and downwards bias scans.

A region of low injection currents (<10 A/cm$^2$) is also measured in reverse bias operation as shown in Fig. 2(a). However, a transition to a higher current injection regime can be seen after the reverse threshold voltage $V_{TH}$=−3.6 V, after which, the current increases monotonically, reaching a current level of ~15 kA/cm$^2$ at −4.4 V. This behavior contrasts with the ~16 V required to achieve similar current levels in the forward direction. This asymmetry in I-V characteristics is a result of the polarization electric fields present in the heterostructure; this interesting connection is discussed in the next section.

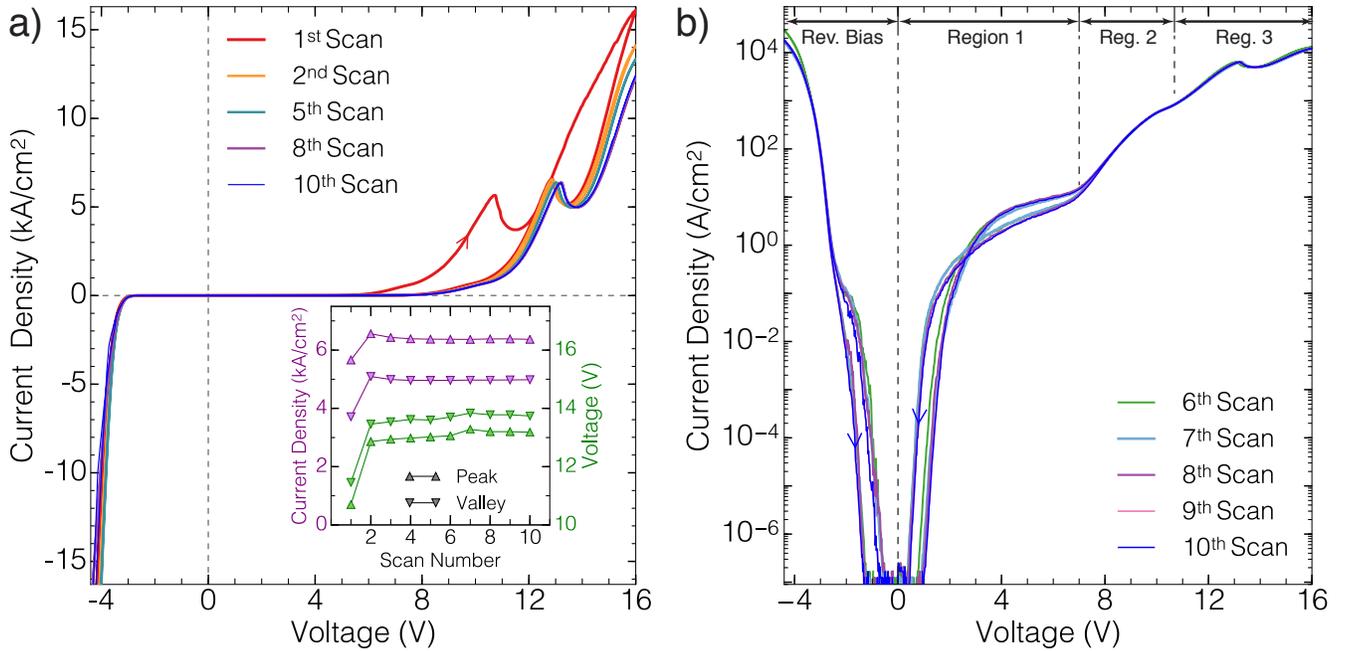

FIG. 3. (a) Subsequent closed-loop bias scans measured at room temperature on the same RTD device. The hysteresis loop measured in the first scan does not appear in the following scans. The onset of NDC stabilizes at ~13.2V with a stable and repeatable peak current density of ~6.4 kA/cm$^2$ as reported in the inset of (a). The stable RTD characteristics confirm the presence of resonant tunneling transport at room temperature with a maximum negative differential conductance of ~ −5.9 kS/cm$^2$. (b) Semilogarithmic plot showing regions with different transport regimes and the presence of a repeatable hysteretic loop under low bias conditions in both directions of operation.

Subsequent closed-loop bias scans were performed in the same device as shown in Fig 3(a). It was found that the resonant bias increases slightly with additional scans and eventually stabilizes at ~13.2 V (inset of Fig. 3(a)). The peak-to-valley current ratio (PVCR) decreases from its initial value of ~1.5 during the first scan to a stable value of ~1.3 in the eighth and subsequent scans. This trend is caused by the larger increase of the valley current which experiences a ~25.5% increment with respect to its initial value, whereas the peak current increases only ~11.1%. This behavior suggests the presence of defects in the AlN barriers; these imperfections act as current leakage paths and degrade the energy filtering function of the tunneling barriers. However, these leakage paths exhibit limited effects and do not prevent resonant tunneling transport of carriers across the heterostructure.

Fig. 3(b) presents the subsequent stable I-V curves in which different conduction regimes can be identified. Low injection currents are measured within region 1, which are a result of the electron blocking effect introduced by the AlN barriers. A repeatable current-voltage hysteresis loop is also measured within this region; this bi-stability might be caused by the charge and discharge of trapping levels within the heterostructure. As the bias is increased, the resonant levels of the GaN quantum well shift towards lower energies and get closer to the Fermi level of the emitter region. The energy alignment between the ground-state of the well and the quasi-continuous scattering states present in the emitter region give rise to an increment in the conductivity. This conduction enhancement starts at ~7 V and extends up to ~10.7 V, constituting the main transport mechanism within region 2. At ~10.7 V, a transition is observed, evidenced by the conductance behavior that initially decreases slightly and then increases again. This modulation is explained by the onset of out-of-resonance conditions between the quasi-continuous scattering states of the emitter and the ground state of the well. Studies done in GaAs/AlGaAs RTDs have also reported the presence of this feature in the rising side of the main resonant peak[29] and this is first time that this behavior is reported in nitride RTDs. As the voltage is further increased to values larger than 10.7 V, the coupling between the accumulation subband (bound state situated below the quasi-continuous scattering states of the emitter, see Fig. 2(c)) and the ground resonant level enhances electron transport across the heterostructure. Eventually, resonant conditions are achieved at ~13.2 V just before the onset of the NDC region, which exhibits a minimum differential conductance of ~ −5.9 kS/cm$^2$. The repeatability of the resonant peak and NDC was tested by the additional closed-loop bias scans, which are also presented in Fig 3(b).

## GaN/AlN RTD Model and Discussion

An electrostatic model has been developed by calculating the distribution of charge, electric field and conduction band energy along the tunneling direction as depicted in Fig. 4(a) and (b). The spontaneous polarization present in nitride materials, as well as the piezoelectric contribution from the strained atomic layers at the GaN/AlN interfaces have been also considered. Sheets of effective polarization charge with charge densities ±$\sigma_\pi$ are included at each of the interfaces as can be seen in the charge diagrams of Fig. 4. The magnitude of the polarization fields generated by these interface charges have been extracted by instersubband absorption measurements, yielding a value of $F_\pi \sim 10$ MV/cm[30].

Under equilibrium conditions, a depletion region builds up in the collector side while an accumulation well is induced next to the emitter barrier. The depletion width $x_d$ and the accumulation charge, modeled by an electron sheet density $n_s$, are both bias dependent. Thus, the electric field $F_0$ generated by the collector space charge is also bias dependent and will be constant within the spacer region where no mobile charge is available. The polarization fields $F_\pi$ generated by the interface sheet charges flip the sign of the electric field inside the barriers— $F_0 - F_\pi$ is negative —while the electric field inside the GaN quantum well remains $F_0$. When a bias is applied to the heterostructure, the space charge regions are modulated and the magnitude of the electric field $F_0$ is in turn modified. The expression for the bias applied to the heterostructure will be:

$$V_{Bias} = \frac{\varepsilon_s}{2eN_d} F_0^2 + (t_s + 2t_b + t_w + t_c)F_0 - 2t_b F_\pi \quad (1)$$

where $\varepsilon_s$ is the GaN dielectric constant; $N_d$, the dopant concentration in the collector side and $e$ is the electron charge. All the thicknesses are as shown in Fig. 4 and $t_c$ is the centroid of the accumulation layer. Using a Schrödinger-Poisson solver[31], we have found that $t_c \approx 1 nm$ under equilibrium conditions and we will consider this value to be approximately constant. Thus, given a voltage bias, Eqn.1 can be solved for $F_0$ as a function of bias, since it is the sole unknown.

Under forward bias operation, $F_0$ increases due to the larger amount of charge present in the emitter accumulation and collector depletion regions. As a consequence, the electric field inside the well also increases and the quantum confined Stark effect generates a decrease in the energies of the bound-states. Resonant conditions arise when the well eigen-energies align with the bottom of the conduction band in the emitter side (red curve in Fig. 4(a)). The energies of the bound-states can be calculated within the framework of a finite quantum well perturbed by the electric field $F_0$ present in the well. Thus if we label the unperturbed ground-state energy as $E_0$, solving for the electric field that satisfies the resonant condition gives

$$F_0^{RES} = \frac{E_0 + F_\pi e t_b}{e\left(\frac{t_w}{2} + t_b + t_c\right)} \quad (2)$$

Using this resonant electric field $F_0^{RES}$ in Eqn. 1 gives the expected theoretical resonant bias.

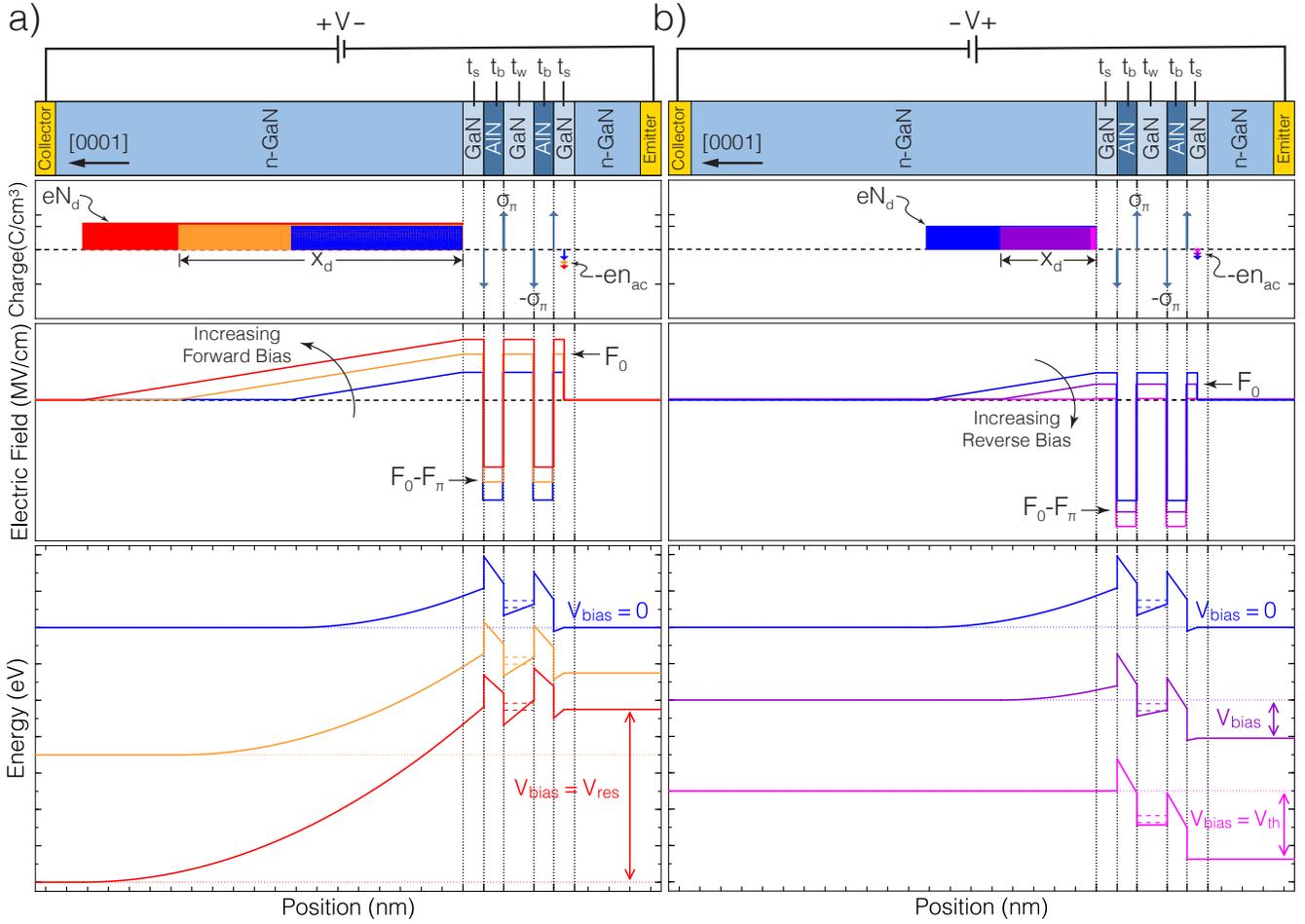

FIG. 4. Distribution of space charge, polarization charge, electric field, and electron energy in the GaN/AlN double barrier heterostructure under forward and reverse bias. The polarity of the applied bias is that of the collector contact with the emitter as ground. (a) Under forward bias, the depletion region in the collector side becomes wider, thus increasing the electric field $F_0$ inside the quantum well and decreasing the magnitude of the negative electric field inside the barriers. The energies of the quantum well bound-states shift towards lower energies as result of the quantum confined Stark effect. Resonant conditions are achieved when the energy of the ground-state bound level aligns with the bottom of the conduction band in the emitter. (b) Under reverse bias operation, the collector depletion region becomes narrower and the quantum well electric field tends to zero while the electric fields inside the barriers increase their magnitude. With a narrower depletion layer, the electrons supporting the reverse current require less energy to reach the collector barrier and tunnel into the emitter.

Furthermore, under reverse bias operation conditions, the electric field $F_0$ will exhibit a decreasing trend as shown in Fig. 4(b). This is a result of the narrowing of the depletion layer in the collector region. As the reverse bias is increased, the electric field within the GaN layers tends to zero and the magnitude of the electric field inside the AlN barriers increases. In addition, as the depletion width narrows, the effective barrier seen by the conduction electrons on the collector side is reduced and the reverse current increases. A critical condition is achieved when $F_0 = 0$ and the depletion region is completely eliminated. Under these flat-band conditions, all the applied voltage is dropped within the barriers and the electric field within them is exactly the polarization field $F_\pi$ (magenta curve in Fig. 4(b)). Thus for an RTD with symmetric barriers, we will have this critical threshold voltage under reverse bias given by the simple expression:

$$V_{TH} = -2t_b F_\pi \tag{3}$$

Fig. 4(b) shows the energy diagram when the reverse threshold bias is applied. Under these conditions, electrons tunneling through the collector AlN barrier will support the overall reverse current. Since the collector barrier is under a high electric field ($F_\pi$), Fowler-Nordheim tunneling is expected to be the main conduction mechanism for voltages above the threshold voltage. This transport regime contrasts with the resonant tunneling transport mechanism supporting the current in forward bias and is a direct consequence of the polarization fields present in the heterostructure. The effects of the internal electric fields on the I-V curves of nitride RTDs have been previously calculated using the transfer matrix formalism[32]. Sakr et al. found that an asymmetric I-V curve is expected in GaN/AlGaN RTDs, constituting a major difference between the operation of III-nitride and other III-V double barrier heterostructures.

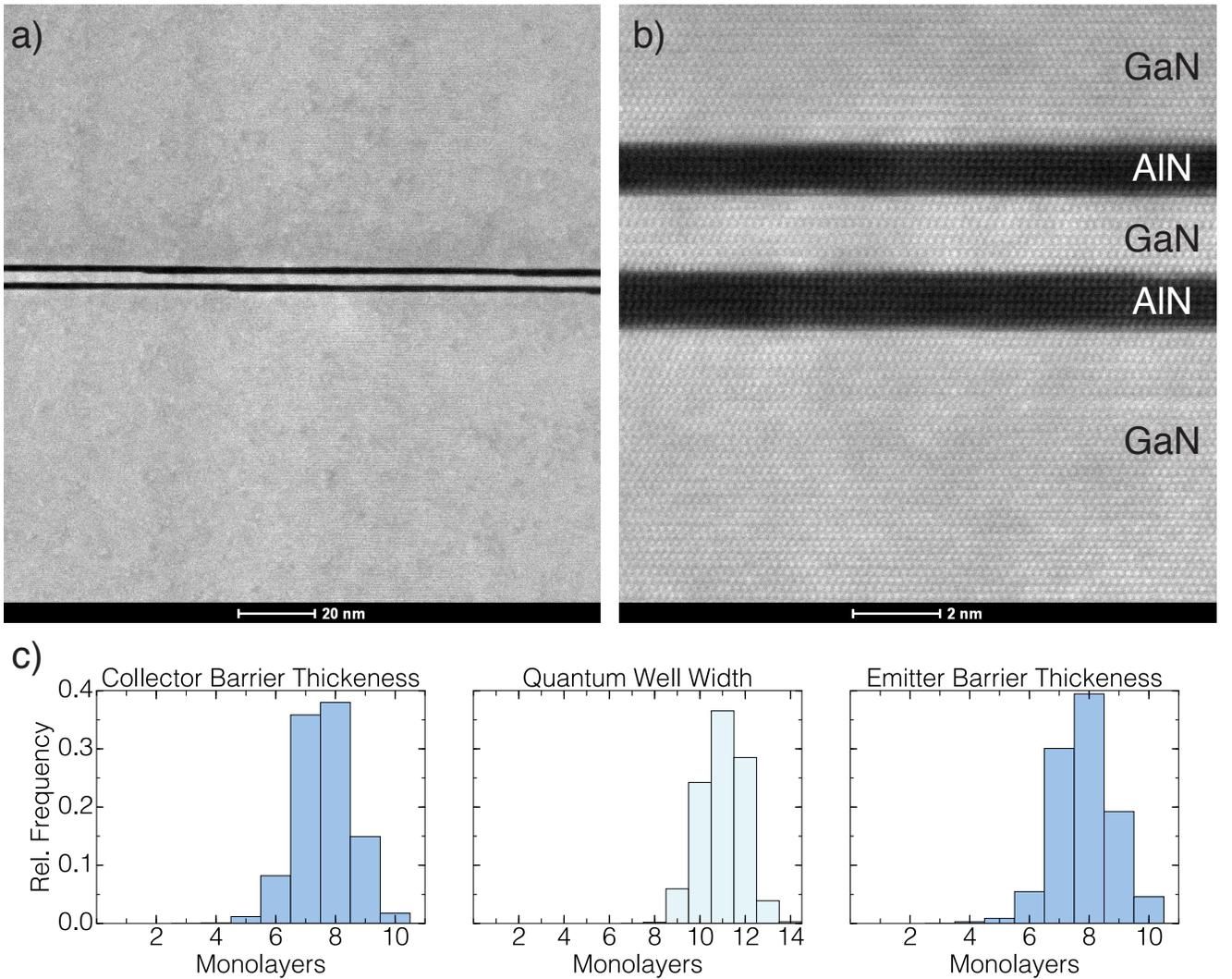

FIG. 5. Scanning transmission electron microscope (STEM) images of the double barrier GaN/AlN RTD heterostructure. (a) A scan over a large area shows the presence of fluctuations in the thicknesses of the ultrathin layers. (b) A small area scan shows the sharp transitions between the AlN barriers and the GaN quantum well. (c) Statistical distribution of the thicknesses of the barriers and the width of the quantum well obtained by digital image processing of Fig. 5(a).

In order to further characterize the grown heterostructures, scanning transmission electron microscopy (STEM) has been employed to measure the thicknesses of the ultra thin layers. Figure 5(a) shows a cross sectional STEM image of the double barriers extending over a distance of ~160 nm perpendicular to the growth direction. It can be seen that fluctuations in the thicknesses of the barriers and well width are present in the heterostructure. The higher magnification image presented in Fig. 5(b) shows the detail of the sharp interfaces over a smaller scanned area. Digital image processing techniques have been applied to Fig. 5(a) to quantify the fluctuations in the thicknesses of the thin nitride layers. The statistical distribution is shown in Fig. 5(c); it was found that both barriers have a mean thickness of 8 MLs (~2 nm) with 1 ML (~0.25 nm) fluctuations. The width of the well, however, presents a mean value of 11 MLs (~2.75 nm), which is 1 ML less than the expected quantum well width. These variations can be incorporated in the electrostatic model so that the theoretical predictions are more accurate. It is found that the expected resonant bias will be $V_{RES} = 10 \pm 1.1V$ including 1 ML fluctuations in the barriers and quantum well thicknesses. The measured experimental value of $V_{RES}^{Exp} \sim 10.7V$ obtained during the first bias scan is within this predicted voltage range. In addition, the threshold voltage has been also calculated and with the inclusion of 1 ML fluctuations in the barrier thicknesses, we found that $V_{TH} = -4 \pm 0.5V$. The experimental value of $V_{TH}^{Exp} \sim -3.6V$ is also within the expected range for the characteristic threshold voltage thus confirming the prediction capability of the electrostatic model.

## CONCLUSIONS

In conclusion, we have experimentally and theoretically investigated the current-voltage characteristics of GaN/AlN double barrier resonant tunneling diodes. The RTD heterostructures were grown by MBE and fabricated into diodes using conventional lithographic techniques. I-V measurements show the negative differential conductance operation at room temperature under forward bias. In contrast, the reverse bias operation is governed by single-barrier tunneling transport and characterized by a well-defined threshold voltage. An electrostatic model is developed, showing that the reverse threshold voltage and forward resonance voltages are intimately linked to the magnitude of the internal polarization fields present in nitride semiconductors. Finally, subsequent measurements show the repeatability of the resonant peak and negative differential conductance, demonstrating that III-nitride RTDs are capable of room temperature resonant tunneling transport. These findings represent a significant step forward in resonant tunneling, intersubband based physics and devices in III-nitrides.